\documentclass[useAMS,usenatbib]{mn2e}

\usepackage{graphicx}

\newcommand{\xmm}{{\it XMM-Newton}}

\newcommand{\kepler}{{\it Kepler}}

\title[Flickering of V1504\,Cyg and V344\,Lyr]{Fast stochastic variability study of two SU\,UMa systems V1504\,Cyg and V344\,Lyr observed by \kepler\ satellite.}
\author[A. Dobrotka, J.-U. Ness and I. Baj\v{c}i\v{c}\'akov\'a]{
A. Dobrotka$^1$\thanks{E-mail: andrej.dobrotka@stuba.sk},
J.-U. Ness$^2$,
I. Baj\v{c}i\v{c}\'akov\'a$^1$\\
$^1$Advanced Technologies Research Institute, Faculty of Materials Science and Technology in Trnava, Slovak University of Technology\\
in Bratislava, Paul\'inska 16, 91724 Trnava, Slovak Republic\\
$^2$XMM-Newton Science Operations Center, European Space Astronomy Center, PO Box 78, 28691 Villanueva de la Ca\~nada, Madrid,\\
}

\begin{document}

\date{Accepted ???. Received ???; in original form \today}

\pagerange{\pageref{firstpage}--\pageref{lastpage}} \pubyear{2016}

\maketitle

\label{firstpage}

\begin{abstract}
We analysed \kepler\ data of two similar dwarf novae V344\,Lyr and V1504\,Cyg in order to study optical fast stochastic variability (flickering) by searching for characteristic break frequencies in their power density spectra. Two different stages of activity were analysed separately, i.e. regular outbursts and quiescence. Both systems show similar behaviour during both activity stages. The quiescent power density spectra show a dominant low break frequency which is also present during outburst with a more or less stable value in V344\,Lyr while it is slightly higher in V1504\,Cyg. The origin of this variability is probably the whole accretion disc. Both outburst power density spectra show additional high frequency components which we interpret as generated by the rebuilt inner disc that was truncated during quiescence. Moreover, V344\,Lyr shows the typical linear rms-flux relation which is strongly deformed by a possible negative superhump variability.
\end{abstract}


\section{Introduction}
\label{introduction}

Accretion processes are generating a large amount of energy. On sub-galactic scales accretion occurs in proto- and T Tauri stars and in binary systems where a primary compact object (a white dwarf, a neutron star or a black hole) is accreting matter from a companion star. The class of X-ray binaries contain a neutron star or black hole, where the companion determines whether a systems belongs to the class of Low-Mass X-ray binaries or High-Mass X-ray binaries (see e.g. \citealt{lewin2010} for a review). All accreting binaries in which the central compact object is a white dwarf belong to the class of Cataclysmic Variables (CV), where some subclasses derive their definition from the nature of the companion, e.g., red giants in Symbiotic Systems (see e.g. \citealt{warner1995}, \citealt{marsh1995} for a review). Strong magnetic fields can influence the mode of accretion, driving the definition of polars (pure accretion stream), intermediate polars (combination of outer accretion stream and inner accretion disk), and non-magnetic CVs in which an accretion disc is present.

We are interested in a specific subclass of non-magnetic CVs called SU\,UMa dwarf novae (DN). In general dwarf novae are alternating between quiescence and regular outbursts. The driving mechanism of this alternation is the viscous-thermal disc instability, generated by variations in ionization state of hydrogen (\citealt{osaki1974}, \citealt{hoshi1979}, \citealt{meyer1981}, \citealt{lasota2001}). During regular outburst the mass accretion rate through the disc is high and the disc is fully developed down to the white dwarf surface, while in quiescence the mass accretion rate is much lower and the disc is truncated, i.e. the inner disc region is missing, and evaporated material is forming an inner hot optically thin geometrically thick X-ray corona (\citealt{meyer1994}). SU\,UMa systems show considerably larger superoutbursts that occur in cycles, in addition to the regular outbursts (see e.g. \citealt{warner1995}, \citealt{lasota2001} for a review) appears in SU\,UMa systems. Two different physical scenarios have been proposed to explain the superoutbursts, i.e. enhanced mass transfer or tidal thermal instability (\citealt{schreiber2004}). The former suggests that the superoutbursts are generated when the disc mass exceeds a critical value, while the latter is based on the outer disc radius expanding to a critical 3:1 resonance radius, where the tidal activity is trigerring the superoutburst. \citet{osaki2013} analysed the same data as we use in this paper, and concluded that the superoutbursts are initiated by a tidal instability.

Every subclass has its own characteristic physical processes and radiation variabilities, but the driving engine is common, i.e. accretion. Therefore, all systems show fast stochastic variability (so-called flickering) as typical manifestation of the underlying accretion process. This variability is observed over a wide range of energies and time scales, from infrared to X-rays and from fractions of a second to tens of minutes (see e.g. \citealt{bruch1992}). The amplitude of this variability is usually defined as square root of the variance and a typical linear relation between this variability amplitude and mean flux with log-normal distribution is observed (\citealt{uttley2005}). This main property of the linear so-called rms-flux relation suggests that the variability is coupled multiplicatively and can be explained by variations in the accretion rate that are produced at different disc radii (\citealt{lyubarskii1997}, \citealt{kotov2001}, \citealt{arevalo2006}). The relation is observed in a variety of accreting systems as X-ray binaries or active galactic nuclei (\citealt{uttley2005}), CVs (\citealt{scaringi2012a}, \citealt{vandesande2015}) and symbiotic systems (\citealt{zamanov2015}).

Flickering is a red noise process characterised by decreasing power with increasing frequency. Such power law in log-log space (the so-called power density spectrum - PDS) has one or more components separated by characteristic break frequencies (\citealt{kato2002}, \citealt{baptista2008}, \citealt{dobrotka2014}), after which the red noise slope becomes steeper towards higher frequencies. Long-term observations of a single sky segment available from the \kepler\ mission offers a unique opportunity to study the PDS in unprecedented detail. Such observational strategy allowed detection of four PDS components in the case of the CV MV\,Lyr (\citealt{scaringi2012b}). An almost continuous light curve of a duration of about 1400 days allowed detailed PDS study during different activity stages of the dwarf nova V1504\,Cyg (\citealt{dobrotka2015b}). Both the outburst and quiescence PDSs of V1504\,Cyg show a rather similar high frequency break, while a lower frequency component is apparently variable between the two activity stages.

V344\,Lyr is a dwarf nova similar to V1504\,Cyg which for a long light curve was observed in the short cadence mode by the \kepler\ mission. We analyse those data in quiescence and outbursts separately in order to search for the multicomponent nature of the PDSs and the typical linear rms-flux relation. Motivated by the results we revisit also the V1504\,Cyg data in order to search for similarities because both V1504\,Cyg and V344\,Lyr have similar orbital parameters and belong to the same subclass of SU\,UMa systems. 

\section{Observations}

The NASA \kepler\ mission (\citealt{borucki2010}) offers a unique opportunity to study fast optical variability of a variety of accreting objects in unprecedented detail, thanks to the high quality of the data: low noise, without any perturbation from the atmosphere, and almost continuous\footnote{The satellite performed quarterly $90\deg$ rolls in order to have the solar panels face to the Sun.} coverage during long time with a cadence of 58.8\,s. This allows detailed studies of the power density spectra (PDS) over a wide frequency range.

As a target of timing analysis we chose two CV systems observed by \kepler, i.e. the SU\,UMa DN systems V1504\,Cyg (KIC number 7446357, orbital period of 1.67\,h) and V344\,Lyr (KIC number 7659570, orbital\footnote{The period is thought to be an orbital period, but following \citet{still2010} it can also be a negative superhump generated by a retrograde-precessing accretion disc.} period of 2.1\,h). The light curves of V1504\,Cyg/V344\,Lyr have durations of approximately 1400 days, and consist of quiescent emission, 118/125 regular outbursts and 11/13 superoutbursts\footnote{For a long-term light curves (first 740 days) see Figs.~1 and 2 in \citet{cannizzo2012}.}. The average quiescent count rates are 448 and 1983 for V344\,Lyr and V1504\,Cyg, respectively. We are interested in short-term variability during quiescence and regular outbursts.

\begin{figure}
\resizebox{\hsize}{!}{\includegraphics[angle=-90]{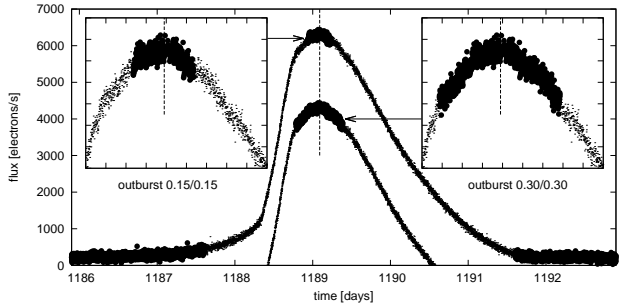}}
\caption{Example of a regular outburst of V344\,Lyr. The dots are the original \kepler\ light curve, and the thick light curve segments are the selected quiescence and outburst subsamples. Two outburst cases are shown, i.e. 0.15/0.15 and 0.30/0.30 (vertically offset downward by a value of 2000 for visualisation purposes) with details in the inset panels. The moment of outburst maximum (taken as maximum of strongly smoothed light curve) is shown by the dashed line.}
\label{lc_outb_details}
\end{figure}

For a quiescent light curve extraction we applied two criteria. First, an upper flux limit\footnote{For V1504\,Cyg in \citet{dobrotka2015b} we used a value of 7000 electrons/s.} of 1200 electrons/s was used to exclude all outburst data above the limit. Second, the times of decline of one outburst and rise into the respective next outburst were determined by a derivative flux limit of 400 electrons/s$^2$. The data in between are the quiescent data. The limits on the derivatives and flux were chosen empirically. Examples of extracted intervals are shown in Fig.~\ref{lc_outb_details}.

The derivative limit method works well for quiescent light curves because the quiescent intervals are long enough and possess high quantity of data, and any border trim does not influence the quality of the resulting PDS. The outbursts are relatively short and the derivative limit method does not work well. Therefore, we adopted an additional method for outburst light curve extraction. We selected a light curve between some time interval before and after the outburst peak\footnote{The outburst peaks are defined as maxima in strongly smoothed outburst light curves.}, taken from a strongly smoothed light curve. As a smoothing we selected averaging of n-successive light curve points. Fig.~2 in \citet{cannizzo2012} suggests that all normal outbursts have a fairly consistent duration, which justifies the use of the same time interval for every outburst. In the subsequent analysis we denote as "outburst 0.10/0.10" a light curve extracted 0.10 days before and 0.10 days after the outburst peak.

\section{Timing analysis}

\subsection{Method}
\label{method}

For the PDS calculation we used the Lomb-Scargle algorithm (\citealt{scargle1982}) which can handle gaps in the light curves or non equidistant data. The low frequency end of the studied PDSs usually depends on the light curve duration. The high frequency end is usually limited by white noise or the power rising to the Nyquist frequency, and we empirically defined this limit to log($f/{\rm Hz}) = -2.2$\footnote{The log is to the base 10.}. The quiescent data were divided into one-day intervals, yielding the low frequency limit approximately log($f/{\rm Hz}) = -5.0$, and every quiescent interval yields an individual PDS. The outburst light curves are too short for further division, therefore we determine a PDS for each full outburst light curve. A mean PDS was calculated from all individual PDSs. We subsequently binned the mean PDS and a mean value with the standard deviation as an indicator of the intrinsic scatter (error estimate) within each averaged frequency bin was calculated from all points in the bin. We used constant number of mean PDS points with a frequency step of log($f/{\rm Hz}) = -5.0$ for every bin to get consistent PDS estimate. This number was chosen empirically (equal to 5) in order to get acceptable signal to noise ratio.

Two y-axis units are commonly used by authors, i.e. a power\footnote{Lomb-Scarge is calculating a normalised power.} $np$ or frequency multiplied by the power $f \times np$. In $np$ units the white noise is a constant, while it is an increasing power law in $f \times np$ units. Furthermore, any red noise is a decreasing power law in $np$ units, while it can be an increasing power law in $f \times np$ units in the case of shallow red noise. Therefore, for visualisation purposes we first use $np$ units, but for subsequent detailed analysis we use $f \times np$ as y-axis units.

All resulting binned PDSs were fitted with individual multicomponent\footnote{$n$-component model is composed of $n$-power laws (linear functions in log-log scale) separated by $n-1$ break frequencies.} models using {\tt GNUPLOT}\footnote{http://www.gnuplot.info/} software, yielding searched break frequencies with standard errors calculated from the variance-covariance matrix directly by the fitting software.

\subsection{V344\,Lyr}

\subsubsection{PDS analysis}

\begin{figure*}
\includegraphics[width=70mm,angle=-90]{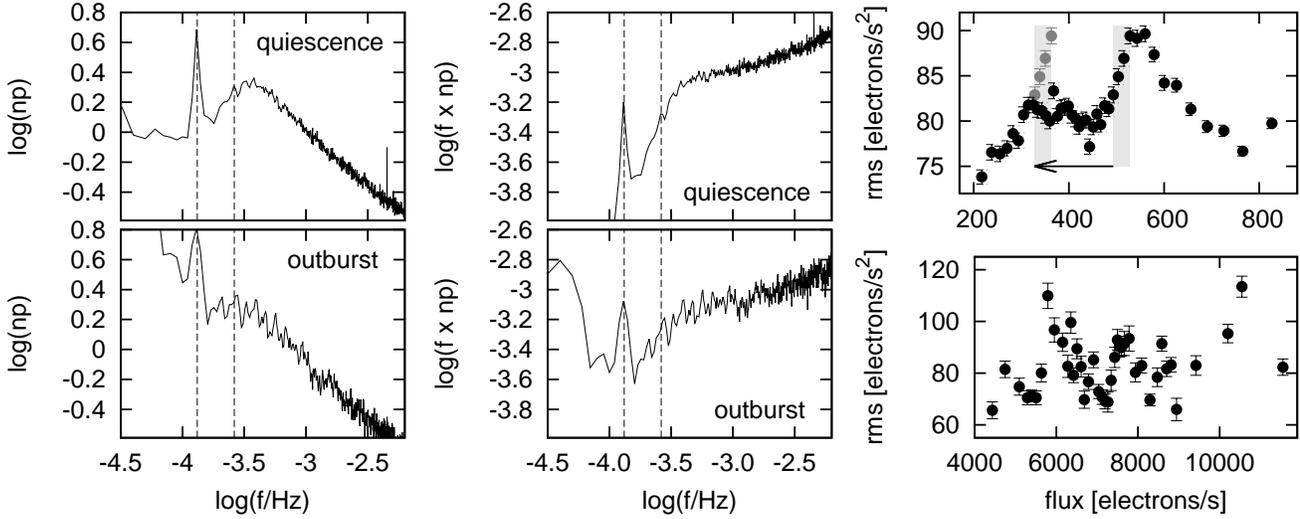}
\caption{Left and middle panels - quiescent and outburst 0.15/0.15 mean PDSs in different y-axis units. The vertical dashed lines indicates the dominant 2.1\,h period with its first harmonic. Right panels - corresponding rms-flux data with error of the mean. The gray data points in the top panel show copied and offset (marked by the arrow) data marked by the shaded area. See text for details.}
\label{pds_all_v344lyr}
\end{figure*}

The mean PDSs of V344\,Lyr calculated from quiescent and outburst data (0.15/0.15 case) in both y-axis units are shown in the left and middle panel of Fig.~\ref{pds_all_v344lyr}. While the middle panel can suggest that the PDS is dominated by the Poissonian noise, the left panel with different y-axis units prove that the variability is a red noise. The difference is due to very low slope of the red noise. Furthermore, the dominant period of 2.1\,h is clear and is marked, together with its first harmonic, by vertical dashed lines. Both PDSs show an obvious break frequency around log($f/{\rm Hz}) = -3.45$. The low frequency part of the outburst PDS is rising because of a long-term trend typical for every outburst. Therefore, for direct comparison of both activity stages we concentrated our study a frequencies above log($f/{\rm Hz}) = -3.8$.

\begin{figure}
\resizebox{\hsize}{!}{\includegraphics[angle=-90]{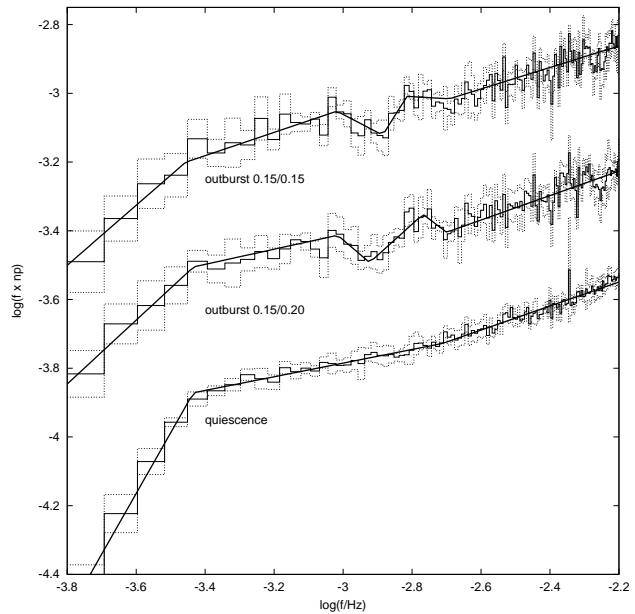}}
\caption{Binned mean PDSs examples of V344\,Lyr for both activity stages with multicomponent fits. Two best outburst cases are shown, i.e. outburst 0.15/0.15 and 0.15/0.20. The latter is shifted vertically by 0.3, while the quiescent PDS by 0.8 for visualisation purposes.}
\label{pds_all_detail_fnp_v344lyr}
\end{figure}

Detailed and binned PDSs from the quiescence is shown in the left panel of Fig.~\ref{pds_all_detail_fnp_v344lyr}. The dominant low break frequency is clear. Using the 2, 3 and 4 component models yield $\chi^2_{\rm red}$ of 1.63, 1.18 and 1.20, respectively. We conservatively assume the scatter within each bin of the averaged PDS to correspond to the statistical errors. The additional break frequency in the best three component models is not the break frequency we are interested in, because the power increases towards higher frequencies after this break. This is a standard shape when red noise is loosing steepness before changing to white noise.

The $\chi^2_{\rm red}$ from different model fitting of various outburst PDSs are summarized in Table~\ref{chi2_fit}. Apparently the shorter the light curve, the poorer the fit. This is not surprising as shorter data sets yield larger PDS scatter and not only the errors are larger, but also any real PDS features are buried in the noise. First we concentrated the fitting to the four-component model. It yields better solutions almost in all cases, at least in those with smallest $\chi^2_{\rm red}$. Just in one case the four-component model did not converged properly. The higher break frequency is even visually non identifiable in this case. A visual inspection suggest an another "peak" at approximately log($f/{\rm Hz}) = -2.8$. Therefore, additional six-component fit were performed with even better $\chi^2_{\rm red}$ (summarized in Table~\ref{chi2_fit}). Two best models with six-component fits are depicted in Fig.~\ref{pds_all_detail_fnp_v344lyr}. Larger extraction limits yielding longer outburst light curves started to loose the higher break frequency because the light curve started to extend too far to the rising or declining branch of the outburst (outburst 0.30/0.30 in Fig.~\ref{lc_outb_details}).
\begin{table}
\caption{$\chi^2_{\rm red}$ values from two (2 pl) and four-component (4 pl) fits for various outburst light curves (lc) for V344\,Lyr (lyr) and V1504\,Cyg (cyg).}
\begin{center}
\begin{tabular}{lcccccr}
\hline
\hline
outburst lc & $\chi^2_{\rm red}$ & $\chi^2_{\rm red}$ & $\chi^2_{\rm red}$ & $\chi^2_{\rm red}$ & $\chi^2_{\rm red}$ & lc duration\\
 & 2 pl & 4 pl & 6 pl & 2 pl & 4 pl & (day)\\
  & lyr & lyr & lyr & cyg & cyg &\\
\hline
0.15/0.20 & 1.70 & 1.37 & 1.26 & 0.76 & 0.77 & 0.35\\
0.10/0.20 & 2.73 & 2.36 & 2.22 & 0.92 & 0.94 & 0.30\\
0.15/0.15 & 1.27 & 1.20 & 1.09 & 1.08 & 1.08 & 0.30\\
0.15/0.10 & 5.19 & 3.73 & 3.33 & 1.74 & 1.69 & 0.25\\
0.10/0.15 & 2.39 & 2.31 & 2.24 & 1.93 & 2.01 & 0.25\\
0.10/0.10 & 3.25 & -- & 2.95 & 2.26 & 2.39 & 0.20\\
0.15/0.05 & 4.44 & 4.49 & 3.04 & 5.47 & 5.51 & 0.20\\
0.10/0.05 & 5.44 & 5.51 & 5.15 & 4.76 & 3.96 & 0.15\\
\hline
\end{tabular}
\end{center}
\label{chi2_fit}
\end{table}

Finally, based on best fits we summarize the resulting break frequencies with standard errors in Table~\ref{pds_param}.
\begin{table*}
\caption{PDS parameters with errors. "npl" means number of power laws used in the fitted model and "dof" means degree of freedom.}
\begin{center}
\begin{tabular}{lccccccr}
\hline
\hline
binary & PDS & log($f_{\rm lower}/{\rm Hz}$) & log($f_{\rm higher,1}/{\rm Hz}$) & log($f_{\rm higher,2}/{\rm Hz}$) & $\chi^2_{\rm red}$ & npl & dof\\
\hline
V344\,Lyr & quiescence & $-3.43 \pm 0.01$ & -- & -- & 1.18 & 3 & 117\\
V344\,Lyr & outb. 0.15/0.15 & $-3.46 \pm 0.13$ & $-3.02 \pm 0.03$ & $-2.81 \pm 0.03$ & 1.09 & 6 & 111\\
V1504\,Cyg & outb. 0.10/0.20 & -- & $-2.97 \pm 0.03$ & & 0.92 & 2 & 69\\
V1504\,Cyg & outb. 0.10/0.20 & $-3.28 \pm 0.11$ & $-2.99 \pm 0.04$ & & 0.94 & 4 & 65\\
V1504\,Cyg & outb. 0.15/0.15 & -- & $-3.03 \pm 0.02$ & & 1.08 & 2 & 69\\
V1504\,Cyg & outb. 0.15/0.15 & $-3.34 \pm 0.07$ & $-3.03 \pm 0.03$ & -- & 1.08 & 4 & 65\\
V1504\,Cyg$^1$ & outb. 0.15/0.20 & $-3.34 \pm 0.03$ & $-3.02 \pm 0.04$ & $-2.87 \pm 0.03$ & 1.03 & 5 & 9\\
V1504\,Cyg$^2$ & quiescence & $-3.38 \pm 0.01$ & -- & & & &\\
\hline
\end{tabular}
\end{center}
\begin{tabular}{l}
$^1$ using the same PDS binning and error estimate as in \citet{dobrotka2015b}\\
$^2$ from \citet{dobrotka2015b}
\end{tabular}
\label{pds_param}
\end{table*}

\subsubsection{Rms-flux}

The absolute rms amplitude of variability is defined as square-root of the variance, i.e.
\begin{equation}
\sigma_{\rm rms} = \sqrt{\frac{1}{N - 1} \sum^N_{i = 1} (\psi_i - \overline{\psi})^2},
\end{equation}
where $N$ is the number of data points, $\psi_i$ is the $i$-th flux point and $\overline{\psi}$ is the mean value of all fluxes $\psi_i$. We subdivided both light curves into small parts each containing 10 points of the light curve. Subsequently for each small part we calculated the corresponding rms and mean flux. The results were averaged using 2000 and 100 rms-flux points for quiescence and outbursts, respectively.

The rms-flux relations of both stages are depicted in the right panels of the Fig.~\ref{pds_all_v344lyr}. The outburst case behavior is not surprising when compared to V1504\,Cyg (\citealt{dobrotka2015b}), but the quiescent case is a surprise. There is a tendency for two linear trends separated by a noisy/variable plateau approximately between flux of 320 and 470 electrons. The high flux tail is totally non standard, i.e. decreasing rms with the increasing flux. The last data point suggests a possible inversion of the trend.

\subsection{V1504\,Cyg}

The V344\,Lyr finding in the previous section motivated us to reanalyse the data of V1504\,Cyg, in order to test whether the low frequency in V1504\,Cyg PDS is really significantly different between the activity stages as presented in \citet{dobrotka2015b}, or the quiescent signal at log($f/{\rm Hz}) = -3.38$ is hidden and non-detected in the outburst PDS noise in \citet{dobrotka2015b}.

We followed the same analysis procedure as in V344\,Lyr case described in Section~\ref{method}, while for the low frequency PDS end we used empirically chosen value of log($f/{\rm Hz}) = -3.45$, because we are interested in the same frequencies as studied in \citet{dobrotka2015b}, i.e. higher than the first harmonic of the orbital frequency. We tried again more outburst variants and fitted models. The situation is not that clear as in V344\,Lyr analysis, i.e. the four-component model does not give clearly better $\chi^2_{\rm red}$, but neither significantly worse. In Table~\ref{chi2_fit} we summarize all $\chi^2_{\rm red}$ from individual fitted models. Apparently the shorter the light curve, the poorer the fit as already seen in V344\,Lyr. Another remark is the $\chi^2_{\rm red}$ difference between two and four-component models for best fits with $\chi^2_{\rm red}$ close to one, they do not differ significantly or are even equal. Therefore, we can not definitely conclude whether the lower break frequency seen in quiescence is a real PDS feature, neither to rule it out. Fig.~\ref{pds_all_detail_fnp_v344lyr} shows the three best cases with both two and four-component fits and the fitted break frequencies from models with $\chi^2_{\rm red}$ closest to one are summarized in Table~\ref{pds_param}.
\begin{figure*}
\includegraphics[width=70mm,angle=-90]{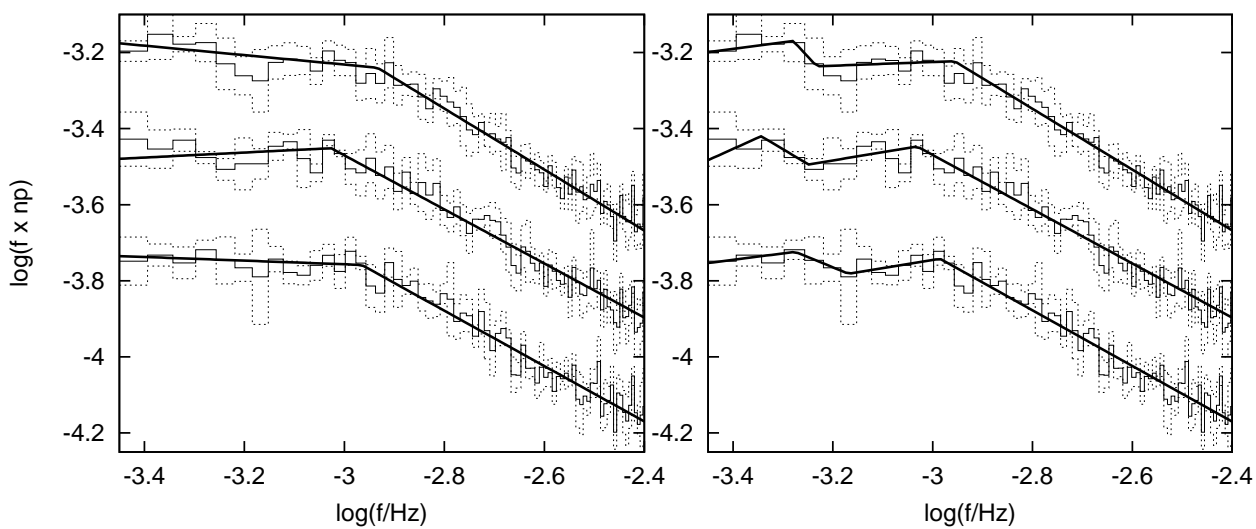}
\caption{V1504\,Cyg outburst (from top to bottom) 0.15/0.20, 0.15/0.15 and 0.10/0.20 PDSs (thin solid line) with estimated errors (dotted line). The thick solid line is the fitted two (left panel) and four-component (right panel) model.}
\label{pds_all_detail_fnp_v1504cyg}
\end{figure*}

\subsection{PDS binning}

In order to test the reality of detected PDS features, we performed an additional analysis of the best PDSs with frequency binning and error estimate as in \citet{dobrotka2015b}. Instead of a constant frequency step, we used a constant logarithmic interval in frequency, and instead of the standard deviation we used error of the mean. The former better describes the intrinsic scatter of the binned mean PDS, while the latter better represents the scatter of the binned points.

\begin{figure*}
\resizebox{\hsize}{!}{\includegraphics[angle=-90]{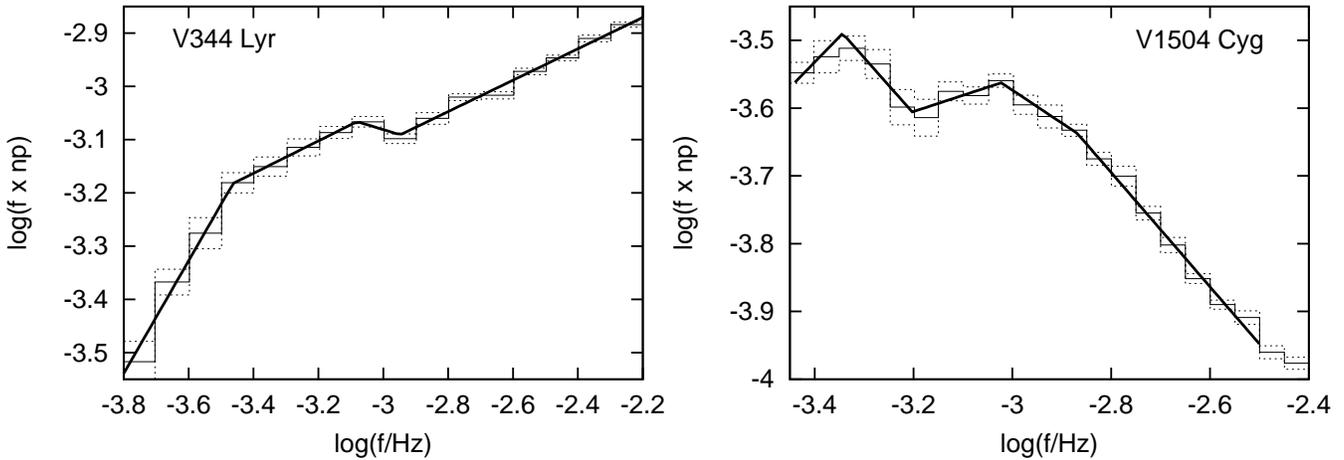}}
\caption{Binned best mean PDSs with different binning procedure than used previously. The errors are the errors of the mean, instead of the standard deviation (see text for details). Left panel - best V344\,Lyr outburst 0.15/0.15 case with a four-component fit. Right panel - best V1504\,Cyg outburst 0.15/0.20 case with a five-component fit.}
\label{pds_detail_fnp_diffbin}
\end{figure*}

The V344\,Lyr case with mean PDS binned into 0.1dex in the logarithmic frequency scale is depicted in the left panel of Fig.~\ref{pds_detail_fnp_diffbin} (outburst 0.15/0.15). The PDS shape is well visible with minimal noise and the four-component nature is unambiguous. Two and three component fits yield unacceptable $\chi^2_{\rm red}$ of 4.36 and 2.73, respectively, while the four-component model yields a value of 1.30. However, this binning version cleanup the "peak" at log($f/{\rm Hz}) = -2.81$, and a smaller frequency step is required. A binning of 0.05dex yields larger data scatter, which negates the meaning of this binning version, and the resulting $\chi^2_{\rm red}$ are not acceptable for any model. However, once the initial parameters are set up, the optimisation method converges and finds a peak around log($f/{\rm Hz}) = -2.81$, but with a doubtful shape and with unconstrained PDS parameters errors\footnote{The algorithm converged yielding PDS parameters with $\chi^2_{\rm red}$ value, but it crashed during error calculation.}, probably because of low data points to fit (degree of freedom). Therefore, this binning version is not adequate for a quantitative study of the V344\,Lyr PDS.

Another situation appears in the case of V1504\,Cyg, with mean PDS binned into 0.05dex (outburst 0.15/0.20, right panel of Fig.~\ref{pds_detail_fnp_diffbin}). The multicomponent shape is again unambiguous with a four-component model yielding a $\chi^2_{\rm red}$ of 1.35. Visual inspection suggests that an additional break frequency is present above the main break at log($f/{\rm Hz}) = -3.0$. Therefore, we fitted the data with a five-component model and we got a $\chi^2_{\rm red}$ of 1.03. The resulting break frequencies are summarized in Table~\ref{pds_param}. This finding suggests, that the outburst light curve selection method used in this paper is better than the one used in \citet{dobrotka2015b}, where we used the same data, binning and error estimate.

\section{Discussion}
\label{discussion}

\subsection{PDS analysis}

We analysed \kepler\ data of two similar dwarf novae V344\,Lyr and V1504\,Cyg in order to search for characteristic break frequencies in PDS from different activity stages, i.e. regular outbursts and quiescence. The analysis of V344\,Lyr is new, while the V1504\,Cyg data were already analysed for this purpose but differently in \citet{dobrotka2015b} yielding different low frequency PDS ends between outbursts and quiescence which we do not find for V344\,Lyr. Such significantly different behaviour of such similar systems is unlikely, which motivated us to reanalyse the V1504\,Cyg data.

\subsubsection{PDS break frequencies}

Using different y-axis units and different outburst light curve extraction limits we can conclude that quiescent and outburst PDSs of V344\,Lyr show a dominant break frequency at log($f/{\rm Hz}) = -3.4$, and a two additional breaks frequencies are present during outburst at log($f/{\rm Hz}) = -3.02$ and log($f/{\rm Hz}) = -2.81$.

V1504\,Cyg data yields similar results for the low frequency break at log($f/{\rm Hz}) = -3.3$ during outburst, which agrees within the estimated errors with a value of log($f/{\rm Hz}) = -3.38$ present during quiescence detected by \citet{dobrotka2015b}. The higher break frequency, previously thought as a modified log($f/{\rm Hz}) = -3.38$ quiescent break frequency, is clear at log($f/{\rm Hz}) = -3.0$ as already detected in outburst by \citet{dobrotka2015b}. Furthermore, an additional break frequency at log($f/{\rm Hz}) = -2.87$ is present during the outburst.

\begin{figure*}
\includegraphics[width=70mm,angle=-90]{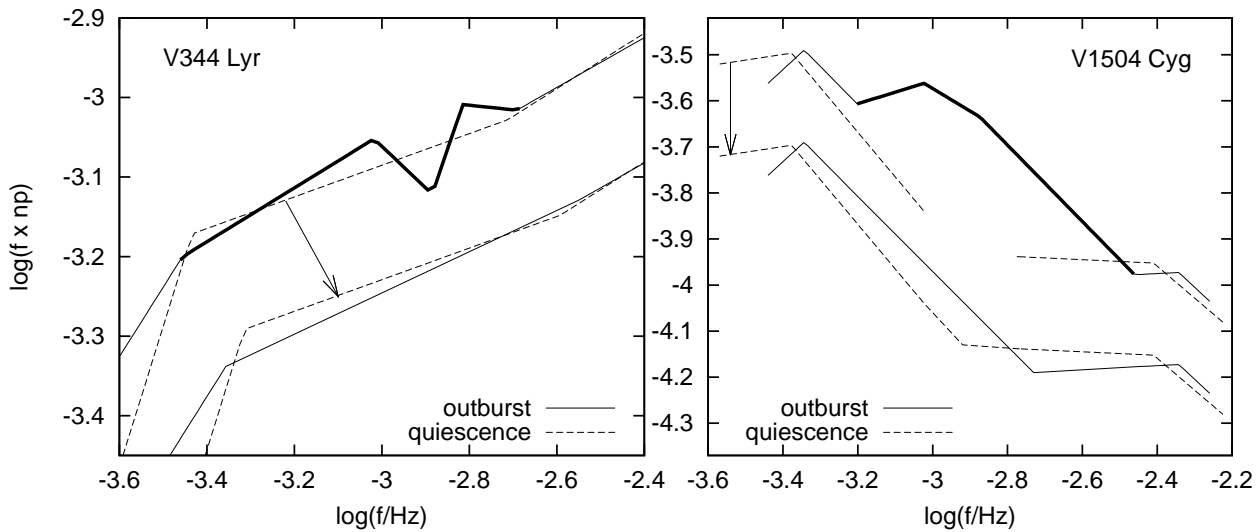}
\caption{Comparison of quiescent and outburst PDS fits of V344\,Lyr (left panel) and V1504\,Cyg (right panel). The upper PDSs with a thick lines are the fits from Fig.~\ref{pds_all_detail_fnp_v344lyr} (outburst 0.15/0.15) and \ref{pds_all_detail_fnp_v1504cyg} (outburst 0.10/0.20). The V1504\,Cyg fit is expanded by two additional power laws to reach also the high frequency part for direct comparison with quiescent fits taken from Fig.~2 of \citet{dobrotka2015b}. The quiescent fits are shifted vertically by -0.10 for V344\,Lyr and -0.22 for V1504\,Cyg to be better comparable with the outburst PDSs. The thick line represents the additional outburst component not detected in quiescence. The lower PDSs are the same situation but shifted (marked by the arrow) for visualisation purposes, and with the thick lines removed to show the similarity of both PDS fits without the additional outburst component.}
\label{pds_fit_comparison}
\end{figure*}

Therefore, both similar systems do show a similar behaviour with a low frequency present during quiescence and an additional PDS complex consisting of two higher break frequencies during outburst. The only significant difference between both systems is the red noise slope which is much shallower in V344\,Lyr. This is shown in Fig.~\ref{pds_fit_comparison} where the PDS fits in both stages are compared.

The V344\,Lyr case is shown in the left panel. The slopes below the lower break frequency are different between quiescence and outburst which is expected because of long-term trend causing shallower slope in outburst (Fig.~\ref{pds_all_v344lyr}). The high frequency ends of both fits have similar slopes. The only significant difference is the PDS pattern characterised by two break frequencies at log($f/{\rm Hz}) = -3.02$ and log($f/{\rm Hz}) = -2.81$ in outburst. This appears like additional components are present in the outburst PDS (marked as thick line) compared to the quiescent case. Otherwise both fits/PDSs have very similar behaviour, which is shown by the shifted PDS (for better comparison) fits where we artificially excluded the additional outburst component\footnote{The remaining PDS data are simply connected with a straight line.}.

The V1504\,Cyg case is depicted in the right panel of Fig.~\ref{pds_fit_comparison}. We used quiescent PDS fits from \citet{dobrotka2015b} and expanded the outburst fit derived in this paper to higher frequencies by adding two more power laws in order to be better comparable with quiescence. The additional outburst PDS components are again marked with thick lines, and removed\footnote{The remaining power laws were simply extrapolated by a straight line.} from the offset PDS for direct comparison with the original one.

This suggests that the quiescent flickering sources remain radiating during the outburst, and an additional flickering source/sources is/are emerging or becoming more pronounced (not detectable during the quiescence) during the outburst.

\subsection{Rms-flux analysis}

In the outburst rms-flux relation of V344\,Lyr we can definitely rule out the typical linearity of the relation as expected from the similar system V1504\,Cyg (\citealt{dobrotka2015b}). The quiescent case is too complicated to confirm or rule out the typical linear relation. Two linear trends marked as shaded areas in Fig.~\ref{pds_all_v344lyr} are clear. The light curve of V344\,Lyr shows a strong signal around 2.1\,h, which may be the orbital period, although following \citet{still2010} this signal can also be a negative superhump generated by the retrograde-precessing accretion disc. Such superhumps can deform the otherwise present linear rms-flux relation by adding some flux offset proportional to the superhump amplitude. This scenario is marked by the arrow in Fig.~\ref{pds_all_v344lyr}, i.e. removing the rms plateau between the flux of 320 and 470 electrons we get the searched linear relation. The grey data points are a copy of the data in the right shaded area and offset horizontally by a flux of 165. We estimated this offset visually and it agrees well with a mean amplitude of the quiescent variability depicted in Fig.~1 in \citet{still2010}. Therefore, the rms-flux relation of V344\,Lyr can still have the typical linear shape, but is somehow deformed due to other variability sources not connected with the accretion process or flickering.

\begin{figure*}
\includegraphics[width=70mm,angle=-90]{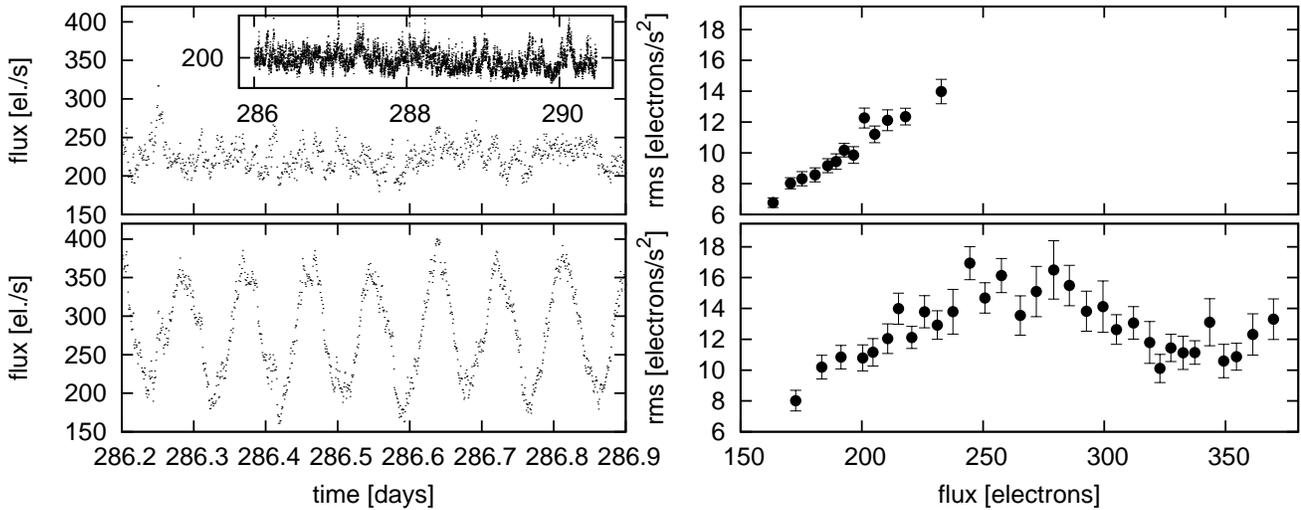}
\caption{Upper left panel - a V1504\,Cyg light curve sequence (all in the inset panel and a detail in the main panel). Lower left panel - the same as upper panel but with added sinusoidal variability. Right panels - corresponding rms-flux data with errors of the mean.}
\label{rms-flux_simul1}
\end{figure*}

We tested this interpretation using a light curve snapshot of V1504\,Cyg where the linear rms-flux relation is present (\citealt{dobrotka2015b}). We selected a light curve subsample with a constant flux level in order to easily transform the original light curve into a "superhumping" light curve using the equation
\begin{equation}
\psi_{\rm new} = 0.1 \times \psi_{\rm orig} + 0.5 \times 165 \times {\rm sin}(t \times 71.4) + 0.5 \times 165,
\end{equation}
where $\psi_{\rm new}$ is the transformed flux, $0.1 \times \psi_{\rm orig}$ is the transformation of the original V1504\,Cyg flux to get similar fluxes as in the left inset panel of Fig.~1 in \citet{still2010}, $0.5 \times 165$ is the estimated half amplitude of the superhumps, $t$ is time coordinate in days and $t \times 71.4$ is a term to get the same "superhump" period of 2.1\,h as in Fig.~1 of \citet{still2010}. The flux transformed light curves with and without the "superhumps" are depicted in the left panels of Fig.~\ref{rms-flux_simul1} with corresponding rms-flux data in the right panels. Clearly the added sine signal is transforming the original linear rms-flux relation into a rising part with an approximately linear trend followed by a rms plateau. Slight decrease of rms toward the end of the plateau is clear and the rms starts to rise again.

\begin{figure}
\includegraphics[width=43mm,angle=-90]{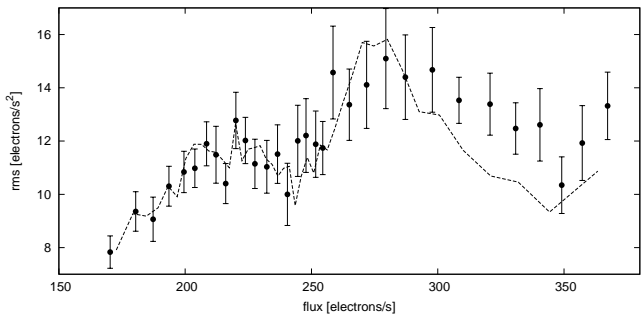}
\caption{Comparison of observed and "simulated" rms-flux data for V344\,Lyr. The data points are calculated from a modified (see text for details) segment of V1504\,Lyr light curve, which exhibits the linear rms-flux relation. The line is the observed rms-flux data from right bottom panel of Fig.~\ref{pds_all_v344lyr} modified horizontally and vertically in order to match the points for direct comparison.}
\label{rms-flux_simul2}
\end{figure}

The real light curve is rather complicated, hence we tried an additional modification. Following \citet{still2010} the superhump variability is not present during whole quiescence. Therefore, the signal is variable and parameters like amplitude can vary or even disappear. We performed the same approach with artificial superhumps, but we divided the light curve into two parts. One is superhumping as in the previous case, while the second half is not. We did not observe any change. Subsequently we tried a variable superhump amplitude with a half amplitude superhump for the second light curve part. The result is depicted in Fig.~\ref{rms-flux_simul2}, where we compare the original observed rms-flux data from Fig.~\ref{pds_all_v344lyr} with the simulated ones. The original V344\,Lyr data were modified horizontally and vertically for direct comparison. Clearly, the behaviour of both is very similar with both rising linear parts, rms plateau and the final decreasing trend with a possible trend inversion at the end. Therefore, we can conclude that V344\,Lyr has the typical linear rms-flux relation, but might be considerably deformed by superhumping variability with variable amplitude.

\subsection{Final model}

In this study we are mainly interested in the difference between quiescence and outburst, because it is well known what is the main difference in dwarf novae, i.e. a truncated vs. fully developed accretion disc (see e.g. \citealt{lasota2001} for a review). Both studied systems show a low frequency break during quiescence which remains present during outburst with the same of slightly higher value. This component should be generated by a structure which is present in both activity stages and is not changing dramatically. The additional and perhaps dominant (in V1504\,Cyg) component is present only during outburst, which suggest that the source is a structure formed during the outburst. Such structure can be the inner disc which is truncated during quiescence. Therefore, we suggest a model, where an accretion disc during quiescence is generating the low frequency break, while it remains present during outburst and the rebuilt inner disc is somehow much more turbulent generating an additional component with a higher break frequency value.

Why should the inner disc be more turbulent than the preexisting truncated disc? The explanation is the Kelvin-Helmholtz instability, occurring whenever there is a velocity gradient to the flow along the direction that separates two fluids. The geometrically thin disc is rotating with a different radial velocity than the inner hot corona surrounding it, which is the source of the mentioned velocity gradient satisfying condition for enhanced turbulence generation of the disc-corona boundary via Kelvin-Helmholtz instability. It is believed, that during quiescence, the corona is formed by evaporation of inner disc material because of inefficient cooling in the low mass accretion rate regime. But it appears that this hot structure is not only present during quiescence. Following \citet{scaringi2014} the highest break frequency of MV\,Lyr in PDS calculated from \kepler\ data is generated by this geometrically thick corona (the so-called sandwiched model) which is confirmed by \xmm\ X-ray observations (Dobrotka et al., in preparation).

The Kelvin-Helmholtz instability as turbulence generator below the corona as a dominant fast variability source is an attractive scenario to explain also the hybrid rms-flux relation observed during outburst in both studied systems. The rms-flux linearity is an imprint of the multiplicative accretion process where variability from different disc rings have characteristic time scales and are correlated, i.e. the mass accretion rate fluctuation at outer disc radii influence the mass accretion rate variability at innermost disc ring. If Kelvin-Helmholtz instability generates turbulences, these appear chaotically and a corresponding mass accretion rate at a certain radius is then not necessarily correlated to a farther-out fluctuations. Therefore, the fast variability generated by the disc satisfying the linear rms-flux relations is complemented by another variability with a different behaviour. \citet{dobrotka2015b} showed that an additional "constant" source deforming the linear rms-flux relation is at work during outburst. It is beyond the scope of this paper to investigate whether the variability generated by Kelvin-Helmholtz instability satisfies this condition\footnote{The "constant" source would mean for V344\,Lyr, that from the point of view of the rms-flux relation, it behaves like a constant source not adding rms (linearly) during flux increases.}.

\begin{figure}
\resizebox{\hsize}{!}{\includegraphics{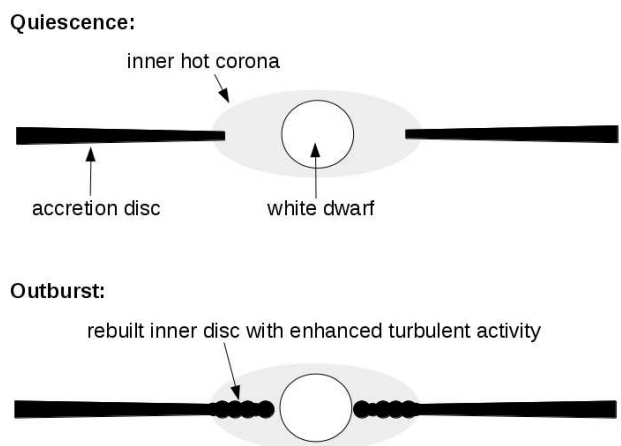}}
\caption{A model of the situation generating a different PDS in quiescence and outburst. See text for details.}
\label{model}
\end{figure}

The proposed final model is depicted in Fig.~\ref{model} together with the inner hot corona as a possible source of an additional high frequency component of V1504\,Cyg between log($f/{\rm Hz}) = -2.3$ and $-2.4$ detected in \citet{dobrotka2015b}.

Having found two characteristic break frequencies in the outburst PDS of both SU\,UMa systems suggests that
the emission from the outburst does not originate from a single structure. More detailed
investigations of structures is beyond the scope of this study.

The main difference between the two systems is the very different slope of the red noise. While more detailed studies of the red noise are beyond the of scope of this work, we find worth emphasizing that V344\,Lyr probably shows superhumps during quiescence (\citealt{still2010}), while they are naturally present only during superoutbursts like in V1504\,Cyg. Following \citet{hellier2001}, the quiescent superhumps are possible in binaries with extreme mass ratios of $q < 0.07$. In such case the angular momentum transport by tidal dissipation is very efficient. Otherwise the turbulent viscosity is the main angular momentum transport mechanism (\citealt{balbus1998}). If there is a significantly different or additional source of angular momentum transport that affects the accretion process, it should influence the variability patterns because flickering is a characteristic manifestation of the underlying turbulent accretion process. Since structures like inner disc and an evaporated corona are similar in both objects, the sources of angular momentum transport are probably different. Therefore, the dominant angular momentum transport mechanisms can influence the red noise slopes (which are different in the two objects), while the accretion structures influence the characteristic break frequencies (which are similar in both objects).

\section{Summary}
\label{summary}

Results of our analysis of optical fast stochastic variability (flickering) of two dwarf novae V344\,Lyr and V1504\,Cyg present in the \kepler\ field can be summarized as follows:

(i) Both systems show a similar PDS shape in quiescence and outburst with similar break frequencies.

(ii) The quiescent PDS is dominated by a single break frequency of approximately log($f/{\rm Hz}) = -3.4$.

(iii) The quiescent break frequency is still present during outburst together with additional components with two higher break frequencies of approximately log($f/{\rm Hz}) = -3.0$ and a second one between $-2.8$ and $-2.9$.

(iv) Rms-flux data of V344\,Lyr do not show the typical linear shape because of possible superhump activity. Taking into account this activity with variable amplitude, a linear trend is present but deformed.

(v) The most likely source of the low frequency component in quiescence and outburst is the accretion disc. The higher frequency component, present only during the outburst, can originates in a re-formed inner disc with somehow enhanced turbulent activity.

\section*{Acknowledgements}

This study was supported by the ERDF - Research and Development Operational Programme under the project "University Scientific Park Campus MTF STU - CAMBO" ITMS: 26220220179. AD acknowledge also the Slovak grant VEGA 1/0335/16. Furthermore, we acknowledge constructive comments of the anonymous reviewer, who helped us to improve this work.

\bibliographystyle{mn2e}
\bibliography{mybib}

\begin{thebibliography}{}

\bibitem[\protect\citeauthoryear{{Ar{\'e}valo} \& {Uttley}}{{Ar{\'e}valo} \&
  {Uttley}}{2006}]{arevalo2006}
{Ar{\'e}valo} P.,  {Uttley} P.,  2006, \mnras, 367, 801

\bibitem[\protect\citeauthoryear{{Balbus} \& {Hawley}}{{Balbus} \&
  {Hawley}}{1998}]{balbus1998}
{Balbus} S.~A.,  {Hawley} J.~F.,  1998, Reviews of Modern Physics, 70, 1

\bibitem[\protect\citeauthoryear{{Baptista} \& {Bortoletto}}{{Baptista} \&
  {Bortoletto}}{2008}]{baptista2008}
{Baptista} R.,  {Bortoletto} A.,  2008, \apj, 676, 1240

\bibitem[\protect\citeauthoryear{{Borucki} et~al.,}{{Borucki}
  et~al.}{2010}]{borucki2010}
{Borucki} W.~J.,  et~al., 2010, Science, 327, 977

\bibitem[\protect\citeauthoryear{{Bruch}}{{Bruch}}{1992}]{bruch1992}
{Bruch} A.,  1992, \aap, 266, 237

\bibitem[\protect\citeauthoryear{{Cannizzo}, {Smale}, {Wood}, {Still} \&
  {Howell}}{{Cannizzo} et~al.}{2012}]{cannizzo2012}
{Cannizzo} J.~K.,  {Smale} A.~P.,  {Wood} M.~A.,  {Still} M.~D.,    {Howell}
  S.~B.,  2012, \apj, 747, 117

\bibitem[\protect\citeauthoryear{{Dobrotka}, {Mineshige} \& {Ness}}{{Dobrotka}
  et~al.}{2014}]{dobrotka2014}
{Dobrotka} A.,  {Mineshige} S.,    {Ness} J.-U.,  2014, \mnras, 438, 1714

\bibitem[\protect\citeauthoryear{{Dobrotka} \& {Ness}}{{Dobrotka} \&
  {Ness}}{2015}]{dobrotka2015b}
{Dobrotka} A.,  {Ness} J.-U.,  2015, \mnras, 451, 2851

\bibitem[\protect\citeauthoryear{{H{\= o}shi}}{{H{\= o}shi}}{1979}]{hoshi1979}
{H{\= o}shi} R.,  1979, Progress of Theoretical Physics, 61, 1307

\bibitem[\protect\citeauthoryear{{Hellier}}{{Hellier}}{2001}]{hellier2001}
{Hellier} C.,  2001, \pasp, 113, 469

\bibitem[\protect\citeauthoryear{{Kato}, {Ishioka} \& {Uemura}}{{Kato}
  et~al.}{2002}]{kato2002}
{Kato} T.,  {Ishioka} R.,    {Uemura} M.,  2002, \pasj, 54, 1033

\bibitem[\protect\citeauthoryear{{Kotov}, {Churazov} \& {Gilfanov}}{{Kotov}
  et~al.}{2001}]{kotov2001}
{Kotov} O.,  {Churazov} E.,    {Gilfanov} M.,  2001, \mnras, 327, 799

\bibitem[\protect\citeauthoryear{{Lasota}}{{Lasota}}{2001}]{lasota2001}
{Lasota} J.,  2001, \nar, 45, 449

\bibitem[\protect\citeauthoryear{{Lewin} \& {van der Klis}}{{Lewin} \& {van der
  Klis}}{2010}]{lewin2010}
{Lewin} W.,  {van der Klis} M.,  2010, {Compact Stellar X-ray Sources}

\bibitem[\protect\citeauthoryear{{Lyubarskii}}{{Lyubarskii}}{1997}]{lyubarskii1997}
{Lyubarskii} Y.~E.,  1997, \mnras, 292, 679

\bibitem[\protect\citeauthoryear{{Marsh}}{{Marsh}}{1995}]{marsh1995}
{Marsh} T.,  1995, The Observatory, 115, 220

\bibitem[\protect\citeauthoryear{{Meyer} \& {Meyer-Hofmeister}}{{Meyer} \&
  {Meyer-Hofmeister}}{1981}]{meyer1981}
{Meyer} F.,  {Meyer-Hofmeister} E.,  1981, \aap, 104, L10

\bibitem[\protect\citeauthoryear{{Meyer} \& {Meyer-Hofmeister}}{{Meyer} \&
  {Meyer-Hofmeister}}{1994}]{meyer1994}
{Meyer} F.,  {Meyer-Hofmeister} E.,  1994, \aap, 288, 175

\bibitem[\protect\citeauthoryear{{Osaki}}{{Osaki}}{1974}]{osaki1974}
{Osaki} Y.,  1974, \pasj, 26, 429

\bibitem[\protect\citeauthoryear{{Osaki} \& {Kato}}{{Osaki} \&
  {Kato}}{2013}]{osaki2013}
{Osaki} Y.,  {Kato} T.,  2013, \pasj, 65, 50

\bibitem[\protect\citeauthoryear{{Scargle}}{{Scargle}}{1982}]{scargle1982}
{Scargle} J.~D.,  1982, \apj, 263, 835

\bibitem[\protect\citeauthoryear{{Scaringi}}{{Scaringi}}{2014}]{scaringi2014}
{Scaringi} S.,  2014, \mnras, 438, 1233

\bibitem[\protect\citeauthoryear{{Scaringi}, {K{\"o}rding}, {Uttley}, {Groot},
  {Knigge}, {Still} \& {Jonker}}{{Scaringi} et~al.}{2012}]{scaringi2012b}
{Scaringi} S.,  {K{\"o}rding} E.,  {Uttley} P.,  {Groot} P.~J.,  {Knigge} C.,
  {Still} M.,    {Jonker} P.,  2012, \mnras, 427, 3396

\bibitem[\protect\citeauthoryear{{Scaringi}, {K{\"o}rding}, {Uttley}, {Knigge},
  {Groot} \& {Still}}{{Scaringi} et~al.}{2012}]{scaringi2012a}
{Scaringi} S.,  {K{\"o}rding} E.,  {Uttley} P.,  {Knigge} C.,  {Groot} P.~J.,
   {Still} M.,  2012, \mnras, 421, 2854

\bibitem[\protect\citeauthoryear{{Schreiber}, {Hameury} \&
  {Lasota}}{{Schreiber} et~al.}{2004}]{schreiber2004}
{Schreiber} M.~R.,  {Hameury} J.-M.,    {Lasota} J.-P.,  2004, \aap, 427, 621

\bibitem[\protect\citeauthoryear{{Still}, {Howell}, {Wood}, {Cannizzo} \&
  {Smale}}{{Still} et~al.}{2010}]{still2010}
{Still} M.,  {Howell} S.~B.,  {Wood} M.~A.,  {Cannizzo} J.~K.,    {Smale}
  A.~P.,  2010, \apjl, 717, L113

\bibitem[\protect\citeauthoryear{{Uttley}, {McHardy} \& {Vaughan}}{{Uttley}
  et~al.}{2005}]{uttley2005}
{Uttley} P.,  {McHardy} I.~M.,    {Vaughan} S.,  2005, \mnras, 359, 345

\bibitem[\protect\citeauthoryear{{Van de Sande}, {Scaringi} \& {Knigge}}{{Van
  de Sande} et~al.}{2015}]{vandesande2015}
{Van de Sande} M.,  {Scaringi} S.,    {Knigge} C.,  2015, ArXiv e-prints

\bibitem[\protect\citeauthoryear{{Warner}}{{Warner}}{1995}]{warner1995}
{Warner} B.,  1995, Cambridge Astrophysics Series, 28

\bibitem[\protect\citeauthoryear{{Zamanov}, {Latev}, {Boeva}, {Sokoloski},
  {Stoyanov}, {Bachev}, {Spassov}, {Nikolov}, {Golev} \& {Ibryamov}}{{Zamanov}
  et~al.}{2015}]{zamanov2015}
{Zamanov} R.,  {Latev} G.,  {Boeva} S.,  {Sokoloski} J.~L.,  {Stoyanov} K.,
  {Bachev} R.,  {Spassov} B.,  {Nikolov} G.,  {Golev} V.,    {Ibryamov} S.,
  2015, \mnras, 450, 3958

\end{thebibliography}

\label{lastpage}

\end{document}